\documentclass{article}
\usepackage{arxiv}

\usepackage{amsmath, amssymb, bm, epsfig}
\usepackage{bm}
\usepackage{bbm}
\usepackage{graphicx}
\usepackage{srcltx}
\usepackage{rotating}
\usepackage{lscape}
\usepackage{booktabs, subfigure}
\usepackage{verbatim, setspace}
\usepackage{calc, color, setspace}
\usepackage[english]{babel}
\usepackage{tabularx}
\usepackage{longtable}
\usepackage{multirow}
\usepackage{booktabs}
\usepackage{ragged2e}
\usepackage{subfigure}
\usepackage{tikz}
\usepackage{ragged2e}
\usepackage{times}
\usepackage{natbib}

\usepackage{url}
\usepackage{rotating}
\usepackage{hyperref}

\newcommand{\app}{\stackrel {{\rm .}}{\sim}}

\newcommand{\ii}{i\in\{1,\ldots,n\}}

\newcommand{\balpha}{\mbox{\boldmath $\alpha$}}

\newcommand{\bmu}{\mbox{\boldmath $\mu$}}

\newcommand{\bnu}{\mbox{\boldmath $\nu$}}

\newcommand{\bSigma}{\mbox{\boldmath $\Sigma$}}
\newcommand{\bepsilon}{\mbox{\boldmath $\epsilon$}}
\newcommand{\bLambda}{\mbox{\boldmath $\Lambda$}}
\newcommand{\bbeta}{\mbox{\boldmath $\beta$}}
\newcommand{\btheta}{\mbox{\boldmath $\theta$}}

\newcommand{\bDelta}{\mbox{\boldmath $\Delta$}}
\newcommand{\bdelta}{\mbox{\boldmath $\delta$}}
\newcommand{\bzeta}{\mbox{\boldmath $\zeta$}}

\newcommand{\SN}{\textrm{SN}}

\newcommand{\SMN}{\textrm{SMN}}
\newcommand{\A}{\mathbf{A}}

\newcommand{\bPsi}{\mbox{\boldmath $\Psi$}}

\newcommand{\bupsilon}{\mbox{\boldmath $\upsilon$}}

\newcommand{\be}{\mathbf{b}}

\newcommand{\yp}{\mathbf{y}}
\newcommand{\xp}{\mathbf{x}}
\newcommand{\y}{\mathbf{y}}
\newcommand{\Y}{\mathbf{Y}}

\newcommand{\bD}{\mathbf{D}}

\newcommand{\blambda}{\mbox{\boldmath $\lambda$}}

\newcommand{\Z}{\mathbf{Z}}

\newcommand{\bW}{\mathbf{W}}

\newcommand{\bH}{\mathbf{H}}
\newcommand{\E}{\textrm{E}}

\newcommand{\SMSN}{\textrm{SMSN}}
\newcommand{\ba}{\mathbf{b}}
\newcommand{\tp}{\mathbf{t}}

\title{A robust nonlinear mixed-effects model for COVID-19 deaths data}

\author{
 Fernanda L. Schumacher \\
  Department of Statistics\\
  Universidade Estadual de Campinas\\
  Campinas, SP Brazil \\
  \texttt{fernandalschumacher@gmail.com} \\
   \And
  Clecio S. Ferreira \\
  Department of Statistics\\
  Universidade Federal de Juiz de Fora\\
  Juiz de Fora, MG Brazil \\
  \texttt{clecio.ferreira@ufjf.edu.br } \\
   \And
 Marcos O. Prates \\
  Department of Statistics\\
  Universidad Federal de Minas Gerais\\
  Minas Gerai, MG, Brazil \\
  \texttt{marcosop@est.ufmg.br} \\
  \And
 Alberto Lachos \\
  National Institute of Neoplastic Diseases\\
  INEN\\
  Lima, Peru \\
  \texttt{alachosd1271@yahoo.com} \\
   \And
 Victor H. Lachos \\ 
  Department of Statistics\\
   University of Connecticut\\
  Storrs, CT 06250 \\
  \texttt{hlachos@uconn.edu} \\
}

\date{June, 26 2020}

\begin{document}
\maketitle
\begin{abstract}
The analysis of complex longitudinal data such as COVID-19 deaths is challenging due to several inherent features: (i) Similarly-shaped profiles with different decay patterns;
(ii) Unexplained variation among repeated measurements within
each country, these repeated measurements may be viewed as clustered data since they
are taken on the same country at roughly the same time; (iii) Skewness, outliers or skew-heavy-tailed noises are possibly embodied within response variables. This article
formulates a robust nonlinear mixed-effects model based in the class of scale mixtures of skew-normal distributions for modeling COVID-19 deaths, which allows the analysts to model such data in the presence of the above described features simultaneously. An
efficient 
EM-type algorithm is proposed to carry out
maximum likelihood estimation of model parameters. The bootstrap method is used to determine inherent characteristics of the nonlinear individual profiles such as confidence interval of the predicted deaths and fitted curves. The target is to model COVID-19 deaths curves from some Latin American countries since this region is the new epicenter of the disease. Moreover, since a mixed-effect framework borrows information from the population-average effects, in our analysis we include some countries from Europe and North America that are in a more advanced stage of their COVID-19 deaths curve.

\end{abstract}

\keywords{COVID-19 deaths data\and Nonlinear mixed-effects models\and  Scale mixtures of skew-normal distributions.}

\section{Introduction}
The world is facing a global pandemic of coronavirus disease (COVID-19), caused by severe acute respiratory syndrome coronavirus 2 (SARS-CoV-2). Initiated at the end of 2019 in Wuhan, China, we are coming, on the 26th of June of 2020, to more than 9 millions cases and 500 thousands deaths spread in 216 countries. The World Health Organization (WHO) has made a joint effort with nations to tackle the disease. Some countries as Germany, Italy, Spain and United Kingdom faced the peak of the disease in March 2020, and now appear to have the disease under control. In other regions such as South-East Asia and Africa 
it seems to be at the beginning (panel at WHO, \url{https://covid19.who.int/}). Currently, Latin American countries are the new epicenter of the disease.

Some institutions around the world have dedicated to data collection and graphical analysis. For example, the panel of the WHO presents graphs of the number of cases and deaths over time for regions and countries. The Johns Hopkins University collects and makes available the daily data, besides producing maps of the occurrence by countries (\url{https://coronavirus.jhu.edu/map.html}). Another available repository is the website Worldometers (\url{https://www.worldometers.info/coronavirus/}). 



There are many institutions in the world modeling and performing forecast of the number of cases and deaths of COVID-19. For example, at the beginning of March 2020, an study of the Oxford University projected more than $450$ thousand of deaths in Brazil (\url{https://www.pnas.org/content/117/18/9696}). 
Another study from the University of Washington's Institute for Health Metrics and Evaluation (IHME) projects a total of $165$ thousand of deaths (\url{https://covid19.healthdata.org/projections}). The Imperial College presented an specific study to Brazil that estimates with high confidence that the reproduction number remained above $1$, indicating that the epidemic was not yet controlled and would continue to grow \citep{ImperialColledge}. In Brazil, there are some organizations studying the evolution of the disease over time allowing online apps with daily updates, for instance, the Department of Statistics of the Federal University of Minas Gerais \citep{covidlp} and the Department of Statistics of the Federal University of Juiz de Fora \citep{JFsalvandotodos}.

The Americas have become the epicenter of the global coronavirus outbreak, logging nearly $4$ million infections and $204,000$ deaths. Brazil, Chile, Colombia, Mexico, and Peru have been particularly hard hit in recent weeks. To point out the possible impact of the disease in Latin America, the \citet{covidlp}, in June $26^{th}$, estimates a total of more than $700$ thousand deaths only for these countries until the end of the pandemic. 

Some studies for modeling the COVID-19 data are based on nonlinear models. \citet{TsallisT2020} proposed a nonlinear model based on volume of stock-markets to predict the number of active cases:
$$N_t=\frac{C(t-t_0)^\alpha}{[1+(q-1)\beta(t-t_0)^\gamma]^{1/(q-1)}},$$
where $C>0$, $\alpha>0$, $\beta>0$, $\gamma>0$, $q>1$ and $t_0 \geq 0$. The constant $t_0$ indicates the first day of appearance of the
epidemic in that particular region; it is conventionally chosen to
be zero for China; for other countries, it is the number of
days elapsed between the appearance of the first case in China
and the first case in that country. The normalizing constant $C$
reflects the total population of that particular country. We refer to \citet{TsallisT2020}, for more details about the interpretation of the parameters $\alpha, \beta, \gamma, q$.

The \citet{covidlp} proposed an hierarchical Bayesian non-Gaussian and non-linear model to capture the dynamics of the epidemic (last accessed on the $26^{th}$ of June on 2020). The daily counts are assumed to come from a Poisson distribution and the daily pandemic evolving by a generalized logistic dynamic:
\begin{eqnarray} \nonumber
 N_t &\sim& Poisson (\mu_t), \\ \label{eq:covidlp}
 \mu_t &=& d \frac{fcae^{-ct}}{(b + e^{-ct} )^{f+1}},
\end{eqnarray}
where  $d$ is responsible to capture subnotification on specific day(s), $c$ control the infection rate, $f$ is an asymmetry parameter, $a$, $b$ and $f$ control the asymptote of the curve, given by $\frac{a}{b^f}$, with the peak occurring at time $t=-\displaystyle\frac{\ln \left(b/f \right)}{c}$. However, to the best of our knowledge, most models developed up to this date are univariate, not taking into account a possible structure of dependence on the disease among countries or regions. Also, as the onset of the disease occurs at different periods between countries, those at a more advanced stage of the disease can provide valuable information to countries at an early stage.

In recent years, nonlinear mixed-effects (NLME) models have been
proposed for modeling many complex longitudinal data \citep{Lindstrombates90, Wu2010}. However, one often assumes that both random error and
random effects are normally distributed, which may not always give
reliable results if the data exhibit excessive skewness and heavy
tailedness, as is the case of COVID-19 deaths data. In this paper, we present a novel class of asymmetric NLME models that provides an efficient estimation of the parameters in the analysis of longitudinal data. We assume that, marginally, the random effects follow a scale mixtures of skew-normal distribution \citep[SMSN,][]{Branco_Dey01} and that the random errors
follow a symmetric scale mixtures of normal distribution \citep[SMN,][]{Lange93} providing an
appealing robust alternative to the usual normal distribution in
NLME models. We propose an approximate likelihood analysis for maximum likelihood (ML) estimation via an EM-type algorithm that produces accurate ML estimates and significantly reduces the numerical difficulty associated with the exact ML estimation. The newly approximate procedures are 
applied to COVID-19 deaths data and it is showed that models with skew-heavy-tailed assumption may provide more reasonable results and these may be important for COVID-19 research in providing quantitative guidance to better
understand the stages and future development of the disease.

The purpose of this study is to predict the number of deaths caused by COVID-19 to short and long term, in countries at early stage such as Peru, Mexico, Chile, Brazil and Colombia along with some countries at a more advanced stage of the disease such as Belgium, Italy, the USA and the United Kingdom. Furthermore, a general bootstrap method is used for constructing 
confidence intervals for the fitted COVID-19 deaths curves and its mode (date of the peak) for the selected countries. The method proposed in this paper is implemented in \texttt{R} \citep{rmanual}, and the codes are available for download from \texttt{Github} (\url{https://github.com/fernandalschumacher/NLMECOVID19}).

The rest of the paper is organized as follows. In Section~\ref{s:data}, we describe the motivating COVID-19 deaths datasets obtained from the John Hopkins repository. In Section~\ref{SECTION3_Model}, we present the methodology and associated ML estimation procedure via the EM-algorithm. In Section~\ref{s:application}, we present our result, where we forecast the total number of deaths in the selected countries for the short term $30$ and $60$ days and long term  $90$ and $150$ days with starting point at June 25$^{th}$ 2020. Finally, the paper concludes with some discussions in Section~\ref{s:conc}.
\section{Motivating COVID-19 data} \label{s:data}


The Johns Hopkins University through the Center for Systems Science and Engineering created a COVID-19 repository (\url{https://github.com/CSSEGISandData/COVID-19/}) that is updated daily with data from many locations of the world. The repository combine data from a variety of official agencies and others reliable sources and unify them.

\begin{figure}[ht]
    \centering
    \includegraphics[width=\textwidth]{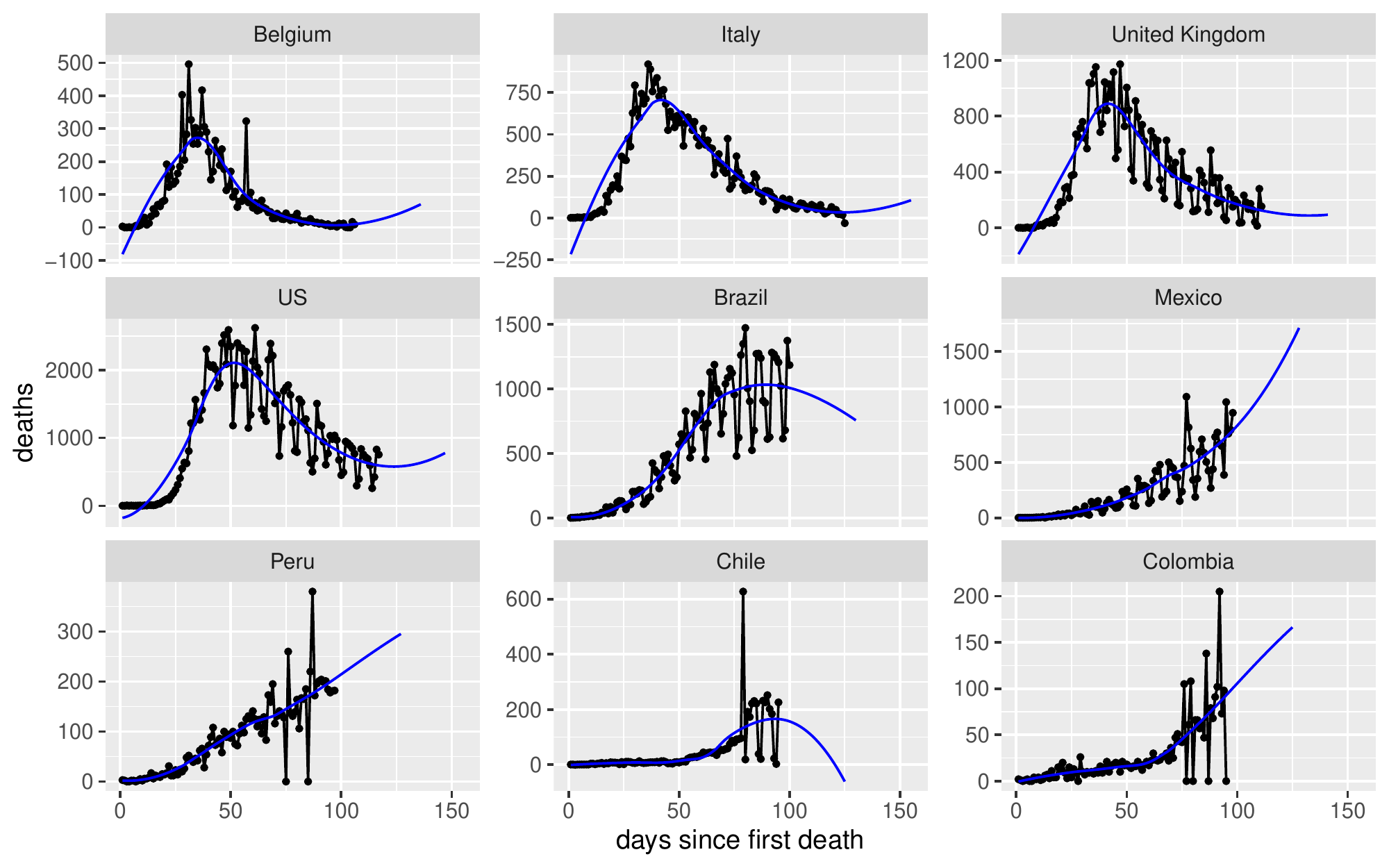}
    \caption{Number of daily reported deaths since first death for the nine countries considered in this study, until 24$^{th}$ of June  2020, with a LOESS non-parametric estimated curve superimposed.}
    \label{fig:data}
\end{figure}

Due to different testing capacity between countries, subnotification is a challenge to understand the true contamination numbers of COVID-19 in the population. Despite that subnotification might also happen in number of deaths, they are less likely to be affected by detection biases. In fact, recent studies use the number of deaths as a proxy measure for COVID-19 cases \citep[see, for example,][]{maugeri2020estimation,ribeiro2020underreporting,amaro2020global}. Figure~\ref{fig:data} shows the reported number of daily deaths in different countries.

As it can be seen the selected countries are in different stages of the COVID-19 pandemic. Belgium, Italy, the United Kingdom (UK), and the United States of America (US) present a controlled number of deaths by the disease. Other countries like Brazil, Mexico, and Peru seem to be around their peak, while Chile and Colombia apparently are in the increasing part of the curve. 
Figure~\ref{fig:data} also presents a LOESS \citep{cleveland1979robust} fit and a $30$ days prediction for each series. Non-parametric curves are alternatives for model fitting because of their flexibility. As it can be seen, the curves have difficulties to capture the rapid increase in deaths in some countries but overall provide a reasonable fit. However, when using this model to make prediction we can clearly see that their extrapolation capacities are not reliable, making necessary the usage of a more adequate statistical parametric model to capture the nature of the pandemic and provide sensible fit and prediction.

Although that Peru moved quickly to lock down its citizens as the pandemic took hold in early March and has extended the lockdown until the end of June 2020, cases nonetheless exploded in May, reaching a peak of more than $8,000$ cases per day late in May, in part explained because the informal employment reaches $73\%$ of the Peruvian labor market. Peru has now reported $268,602$ cases and $8,761$ deaths of COVID-19 \citep{Worldometers}, the second-highest number of confirmed cases of the disease in Latin America, behind Brazil, and the seventh-highest globally. 

In Mexico, the accumulated COVID-19 cases reached $202,951$ and a death total of $25,060$ until June $26^{th}$. Moreover, the statistics show that the curve is ascending \citep{Worldometers}. \citet{TorrealbaCH2020} presented a modeling and prediction of accumulated cases of COVID-19 infection in Mexico using the models of Gompertz and Logistic and a framework of Artificial Neural Network. These models predict the peak between May $8^{th}$ and June $25^{th}$, which might be unrealistic as presented in Section~\ref{s:application}. 

Colombia is in an ascending stage of the curve of cases and deaths, having accumulated $80,599$ cases and $2,654$ deaths until June, $26^{th}$ according to \citet{Worldometers}. \citet{Rivera-Rodriguez2020.04.14.20065466} used a SEIR model to estimate the number of patients that would required Intensive Care Units (ICU) care (critical), and only hospital care (severe) in order to manage their limited resources. \citet{arm2020sir} used a SIR model to estimate the number of cases and deaths in Colombia using data of Italy and South Korea where the peak of infections would be by June $29^{th}$. As shown in Figure~\ref{fig:data}, Colombia presents an increasing pattern and the peak should be observed somewhere in the future.

For Chile, the statistics are of $263,360$ accumulated cases and a total of $5,068$ deaths until June $26^{th}$ \citep{Worldometers}. It is important to observe that there are two days with observations far above the standard:  number of deaths of $649$ (June $8^{th}$) and number of cases of $36,179$ (June $18^{th}$).

Lastly, Brazil presented $1,244,419$ cases and $55,304$ deaths until June $26^{th}$ according to \citet{Worldometers} and is the country with second-highest number of confirmed cases and deaths in the world. \citet{covidlp} estimates that the Brazilian peak of deaths should happen in June $06^{th}$ and a estimated total number of death of $99,932$ which is closely related to the predictions presented in Section~\ref{s:application}. From Figure~\ref{fig:data} we can see that Brazil is the country in more advanced stage of the evolution of the disease in the Latin American region. 

The different stages of spread of the disease are an important information that should be used into modeling. In the next section we introduce our methodology that jointly accommodates the different stages of the diseases and borrow information of the different time series to provide a more robust and reliable fit and prediction.

\section{The model} 
\label{SECTION3_Model}

\subsection{Scale mixtures of skew-normal distributions}
The idea of the SMSN distributions originated from an early work by
\cite{Branco_Dey01}, which included the skew-normal (SN)
distribution as a special case. We say that a $p\times 1$ random
vector $\textbf{Y}$ follows a SN distribution with $p\times 1$
location vector $\bmu$, $p\times p$ positive definite dispersion
matrix $\bSigma$ and $p\times 1$ skewness parameter vector
$\blambda,$ (often known as the shape parameter) and write
$\textbf{Y}\sim SN_p(\bmu,\bSigma,\blambda),$ if its probability
density function (pdf) is given by
\begin{equation}
f(\mathbf{y})= 2{\phi_p(\mathbf{y};\bmu,\bSigma)
\Phi(\blambda^{\top}\bSigma^{-1/2}(\mathbf{y}-\bmu))},\label{denSN}
\end{equation}
where  $\phi_p(.;\bmu,\bSigma)$ stands for the pdf of the
$p$--variate normal distribution with mean vector $\bmu$ and
dispersion matrix $\bSigma$, $N_{\texttt{p}}(\bmu,\bSigma)$ say, and
$\Phi(.)$ is the cumulative distribution function (cdf) of the
standard univariate normal. {Note for $\blambda=\bf 0$,
(\ref{denSN}) reduces to the symmetric $N_p(\bmu,\bSigma)$-pdf,
while for non-zero values of $\blambda$, it produces a perturbed
(asymmetric) family of $N_p(\bmu,\bSigma)$-pdf's.} Let $\Z=\Y-\bmu.$
Since $a\Z\sim SN_p(\textbf{0},a^2\bSigma,\blambda),$ for all scalar
$a>0,$ the SMSN family can be defined as follows: a SMSN distribution
is that of a $p-$dimensional random vector
\begin{equation}
 \Y=\bmu+U^{-1/2}\Z,\label{stoNI}
\end{equation}
where $U$ is a positive random variable with the cdf $H(u;\bnu)$ and
pdf $h(u;\bnu)$, and independent of the
$SN_p(\textbf{0},\bSigma,\blambda)$--random vector $\Z.$ Here $\bnu$
is a scalar or vector parameter indexing the distribution of the
mixing scale factor  $U$. Given $U=u$, $\Y$ follows a multivariate
skew--normal distribution with location vector $\bmu$, scale matrix
$u^{-1}\bSigma$ and skewness parameter vector $\blambda$, i.e.,
$\Y|U=u\sim \SN_p(\bmu,u^{-1}\bSigma,\blambda)$. Thus, by
(\ref{denSN}), the marginal pdf of $\Y$ is
\begin{equation}
f(\mathbf{y})=
2\int^{\infty}_0{\phi_p(\mathbf{y};\bmu,u^{-1}\bSigma)
\Phi(u^{1/2}\blambda^{\top}\bSigma^{-1/2}(\mathbf{y}-\bmu))}dH(u;\bnu).\label{denSNI}
\end{equation}
The notation $\mathbf{Y}\sim \SMSN_p(\bmu,\bSigma,\blambda;H)$ will
be used when $\Y$ has pdf (\ref{denSNI}). The  asymmetrical  class
of SMSN distributions includes the skew--$t$ (ST), the skew--slash
(SSL), and the skew--contaminated normal (SCN). All these
distributions have heavier tails than the skew-normal and can be
used for robust inferences.  When $\blambda=\mathbf{0}$, the SMSN
class reduces to the scale mixtures of normal (SMN) class, which is represented by the pdf
$f_0(\mathbf{y})=\int^{\infty}_0{\phi_p(\mathbf{y};\bmu,u^{-1}\bSigma)
}dH(u;\bnu)$. We use the notation $\mathbf{Y}\sim
\textrm{\SMN}_p(\bmu,\bSigma;H)$ when $\mathbf{Y}$ has distribution
in the SMN class. We refer to \cite{schumacher2020scale} for
details and additional properties related to this class of
distributions.

\subsection{Model formulation}
In this section, we present the models and methods in general forms,
illustrating that our methods may be applicable in other
applications as well. Denote the number of
subjects by $n$ and the number of measurements on the $i$th subject
by $n_i$. For notational convenience, let $x_{ij}$ $(i = 1, 2,\ldots
,n; j = 1, 2, \ldots , n_i )$ be a vector incorporating independent
variables such as number of ICU beds, $\bbeta_{ij} = (\beta_{1ij}, \ldots , \beta_{sij})^{\top}$,
$\bbeta=(\beta_1,\ldots,\beta_r)^{\top} (r>s)$. The NLME model
can be written as
\begin{eqnarray}
\textbf{y}_i&=&\eta_i(t_{ij},\bbeta_{ij})+\bepsilon_i,\,\,\,\bbeta_{ij}=d(x_{ij},\bbeta,\be_i),\label{modeleq11}
\end{eqnarray}
where the subscript $i$ is the subject index, $\yp_i = (y_{i1}, \ldots , y_{in_i} )^{\top}$, with $y_{ij}$ being the response value for
individual $i$ at time $t_{ij}$, $\eta_i(t_{ij} , \bbeta_{ij}) =
(\eta(t_{i1} , \bbeta_{i1})^{\top},\ldots, \eta(t_{in_i} ,
\bbeta_{in_i}))^{\top}$, with $\eta({\cdot})$ being a nonlinear known
function, $\bepsilon = (\epsilon_{i1}, \ldots , \epsilon_{in_i}
)^{\top}$ is random error vector,  $d(.)$ is an $s$-dimensional
linear function, $\be_i = (b_{1i} ,\ldots, b_{qi} )^{\top}$ is the
vector of random effects $(q\leq s)$. 

Following  \cite{schumacher2020approximate}, we assume that
\begin{equation}\label{modSnmis2} \left(\begin{array}{c}
        \mathbf{b}_i \\
        \bepsilon_i
      \end{array}
\right)\buildrel ind\over\sim
\SMSN_{q+n_i}\left(\left(\begin{array}{c}
                               c\bDelta \\
                               \mathbf{0}
                             \end{array}
\right),\left(\begin{array}{cc}
          \mathbf{D} & \mathbf{0} \\
           \mathbf{0} & \sigma^2\mathbf{I}_{n_i}
         \end{array}
\right), \left(\begin{array}{c}
                               \blambda \\
                               \mathbf{0}
                             \end{array}
\right); H \right),\ii,
\end{equation}
where $c=-\sqrt{\frac{2}{\pi}} k_1$, with $k_{1} = E\{U^{-1/2}\},$
$\bDelta=\bD^{1/2}\bdelta$, with
$\bdelta=\blambda/\sqrt{1+\blambda^{\top}\blambda}$, $\sigma^2$ is
the unknown within-subject variance, $\bD=\bD(\balpha)$ is the
$q\times q$ variance-covariance matrix of $\be_i$, which depends on
unknown and reduced parameters $\balpha$ of dimension $v\times 1$
and $\blambda=(\lambda_1,\ldots,\lambda_q)^{\top}$ is a $q\times 1$
vector of skewness parameters for the random effects. Using the
definition of a SMSN random vector and (\ref{modSnmis2}), it follows
that marginally
\begin{equation}
\textbf{b}_i\buildrel iid\over\sim \SMSN_q(c\bDelta
,\mathbf{D},\blambda; H)\quad\textrm{and}\quad\bepsilon_i\buildrel
ind\over\sim \SMN_{n_i}(\mathbf{0},\sigma^2\mathbf{I}_{n_i};
H),\quad\ii,
 \label{modeleq2}
\end{equation}
so that $E\{\be_i\}=E\{\bepsilon_i\}=\mathbf{0}$. Thus this model
considers that the $\bepsilon_i$'s, related to within-subject errors
are symmetrically distributed, while the distribution of  random
effects is assumed to be asymmetric and with mean zero.  In
addition, under this consideration the regression parameters are all
comparable. 

Let
$\btheta=(\bbeta^{\top},\sigma^2,\balpha^{\top},\blambda^{\top},\bnu^\top)^{\top}$,
then classical inference
on the parameter vector $\btheta$ is based on the marginal
distribution for $\mathbf{Y}=(\yp^{\top}_1,\ldots,\yp^{\top}_n)$.
Thus, the integrated likelihood of
(\ref{modeleq11})-(\ref{modSnmis2}) for $\btheta$ in this case is
given by
\begin{eqnarray}
L(\btheta)&=&2\, \prod^n_{i=1}\int^{\infty}_{0}\int_{\mathbb{R}^q}
\phi_{n_i}(\mathbf{y}_i;\eta_i(t_{ij},\bbeta_{ij}),u_i^{-1}\sigma^2
\mathbf{I}_{n_i})\phi_q(\be_i;c\bDelta,u^{-1}_i\mathbf{D})\nonumber\\
&\times&\Phi(u_i^{1/2}\blambda^{\top}\bD^{-1/2}(\be_i-c\bDelta))d\be_idH(u_i;\bnu),\label{cor1eq}
\end{eqnarray}
which generally does not have a closed form expression because the
model function is not linear in the random effects. In the normal
case, to make the numerical optimization of the likelihood function
in a tractable problem, different approximations to (\ref{cor1eq})
have been proposed. Some of these methods consist of taking a
first-order Taylor expansion of the model function around the
conditional models of the random effects $\be$
\citep{Lindstrombates90}. Following this idea, the marginal
distribution of $\mathbf{Y}_i,$ for $\ii,$ can be approximated as
\begin{equation} \label{aproxLK}
\Y_i \app
\SMSN_{n_i}(\eta_i(t_{ij},d(x_{ij},\bbeta,\widetilde{\be}_i))-\widetilde{\bH}_i(\widetilde{\ba}_i-c\bDelta),\widetilde{\bPsi}_i,\widetilde{\bar{\blambda}};H),
\end{equation}
where 
 $\widetilde{\bPsi}_i=\widetilde{\bH}_i \bD
\widetilde{\bH}_i^{\top}+\sigma^2 \mathbf{I}_{n_i}$,
$\widetilde{\bH}_i=\displaystyle\frac{\partial
\eta_i(t_{ij},d(x_{ij},\bbeta,{\be}_i))}{\partial{\ba}^{\top}_i}|_{\ba_i=\widetilde{\ba}_i},$ $\widetilde{\ba}_i$ is an expansion point in a neighborhood of $\ba_i$,
$\widetilde{\bar{\blambda}}_{i}=
\displaystyle\frac{\widetilde{\bPsi}_i^{-1/2}\mathbf{\widetilde{H}}_i\mathbf{D}{\bzeta}}
{\sqrt{1+\bzeta^{\top}\widetilde{\bLambda}_i\bzeta}},$
$\textrm{with}\,\,\, \bzeta=\mathbf{D}^{-1/2}\blambda,
\widetilde{\bLambda}_i=(\mathbf{D}^{-1}+\sigma^{-2}\mathbf{\widetilde{H}}_i^{\top}
\mathbf{\widetilde{H}}_i)^{-1}$, and $`` \app"$ denotes approximated
in distribution.\\

The approximated empirical Bayes estimator of
$\mathbf{b}_i$, denoted by $\widetilde{\mathbf{b}}_i$, obtained by the
conditional mean of $\mathbf{b}_i$ given
${\Y}_i$, is
\begin{eqnarray}\label{MSQ}
{\widetilde{\mathbf{b}}}^{(k)}_i(\btheta)&\approx&\E\{\mathbf{b}_i|{\Y}_i=\yp_i,\btheta\}\approx\widetilde{\bmu}_{bi}+\frac{\widetilde{\tau}_{-1i}}{\sqrt{1+\bzeta^{\top}\widetilde{\bLambda}_i\bzeta}}\,\widetilde{\bLambda}_i\bzeta,
\end{eqnarray}
where
$\widetilde{\bmu}_{bi}=c\bDelta+\mathbf{D}\widetilde{\bH}^{\top}_i\widetilde{\bPsi}^{-1/2}_i\widetilde{\y}_{0i}$
and
$\widetilde{\tau}_{-1i}=\E\{U^{-1/2}W_{\Phi}(U^{1/2}\widetilde{\A}_i)|{\Y}_i\}$,
with $W_{\Phi}(x)=\phi_1(x)/\Phi(x)$, $x\in \mathbb{R}$,
$\widetilde{\y}_{0i}=\bPsi^{-1/2}_i(\yp_i-\eta_i(t_{ij},d(x_{ij},\bbeta,\widetilde{\be}^{(k-1)}_i))+\widetilde{\bH}_i\widetilde{\ba}^{(k-1)}_i-c\widetilde{\bH}_i\bDelta)$
and
$\widetilde{\A}_i=\widetilde{\bar{\blambda}}^{\top}_{i}\widetilde{\y}_{0i}.$ We refer to \cite{Lachos_Ghosh_Arellano_2009}, \cite{schumacher2020scale} and \cite{schumacher2020approximate}, for further details.\\

\subsection{Approximate ML estimation via the EM algorithm}\label{subsec:approxML}

In this section, we demonstrate how to use the EM algorithm \citep{Dempster77} to
obtain approximate maximum likelihood (ML) estimator of a SMSN--NLME model. We denote the current
estimates of $(\bbeta,\ba_i)$ by
$(\widetilde{\bbeta},\widetilde{\ba}_i)$. In this case, the
linearization procedure \citep{Wu2010}
consists of taking the first-order Taylor expansion of $\eta_{i}$
around the current parameter estimate $\widetilde{\bbeta}$ and the
random effect estimates $\widetilde{\ba}_i$, which is equivalent to iteratively solving the
following linear mixed-effect (LME) model
\begin{equation}\label{aprox}
\widetilde{\mathbf{Y}}_i=\widetilde{{\bW}}_i\bbeta+\widetilde{{\bH}}_i\ba_i+\bepsilon_i,
\end{equation}
for $\ii$, where
$\widetilde{\mathbf{Y}}_i=\mathbf{Y}_i-\widetilde{\eta}_i(t_{ij},d(x_{ij},\widetilde{\bbeta},\widetilde{\be}_i)),$
$\widetilde{\eta}_i(t_{ij},d(x_{ij},\widetilde{\bbeta},\widetilde{\be}_i))=\eta_i(t_{ij},d(x_{ij},\bbeta,\widetilde{\be}_i))-\widetilde{\bH}_i\widetilde{\be}_i-\widetilde{\bW}_i\widetilde{\bbeta}$,
$\widetilde{\bPsi}_i=\widetilde{\bH}_i \bD
\widetilde{\bH}_i^{\top}+\sigma^2 \mathbf{I}_{n_i},$
$\widetilde{\bH}_i=\displaystyle\frac{\partial
\eta_i(t_{ij},d(x_{ij},\widetilde{\bbeta},{\be}_i))}{\partial{\ba}^{\top}_i}|_{\ba_i=\widetilde{\ba}_i},$
$\widetilde{\bW}_i=\displaystyle\frac{\partial
\eta_i(t_{ij},d(x_{ij},\bbeta,\widetilde{\be}_i))}{\partial{\bbeta}^{\top}}|_{\bbeta=\widetilde{\bbeta}},$
$\widetilde{\bar{\blambda}}_{i}=\displaystyle\frac{\widetilde{\bPsi}_i^{-1/2}\mathbf{\widetilde{H}}_i\mathbf{D}\bzeta}
{\sqrt{1+\bzeta^{\top}\widetilde{\bLambda}_i\bzeta}},$
$\textrm{with}\,\,\, \bzeta=\mathbf{D}^{-1/2}\blambda,
\widetilde{\bLambda}_i=(\mathbf{D}^{-1}+\sigma^{-2}\mathbf{\widetilde{H}}_i^{\top}
\mathbf{\widetilde{H}}_i)^{-1},$ $ \be_i\buildrel iid\over\sim \SMSN_q(c\bDelta,\mathbf{D},\blambda,
H)$ and $\bepsilon_i\buildrel ind.\over\sim
\SMN_{n_i}(\mathbf{0},\sigma^2 \mathbf{I}_{n_i}H)$. The model defined in (\ref{aprox}) can be seen as a slight modification of the general SMSN-LME model proposed by \cite{Lachos_Ghosh_Arellano_2009} and \cite{schumacher2020scale}, where a simple and efficient EM-type algorithm for iteratively computing ML estimates of the parameters in the SMSN-LME model has been proposed and for which we use the \verb"skewlmm" package in {R} \citep{skewlmm-manual,rmanual}. The approximated likelihood function, derived from (\ref{aproxLK}),
can be easily computed as a byproduct of the EM-algorithm and is used for monitoring convergence and for model selection, such as,  the Akaike (AIC) and Bayesian Information Criterion (BIC) \citep{wit2012all}.

\subsection {Futures observations\label{notes}}

Suppose now that we are interested in the prediction of
$\mathbf{y^+}_i$, a future $\upsilon\times 1$ vector of measurement
of $\mathbf{Y}_i$, given the observed measurement
$\mathbf{Y}=(\mathbf{Y}^{\top}_{(i)},\mathbf{Y}^{\top}_i)^{\top}$,
where
$\mathbf{Y}_{(i)}=(\mathbf{Y}^{\top}_1,\ldots,\mathbf{Y}^{\top}_{i-1},\mathbf{Y}^{\top}_{i+1},\ldots,\mathbf{Y}^{\top}_{n})$.
The minimum mean square error (MSE) predictor of $\y^+_i$, defined as the conditional
expectation $\y_i^{+}$ given $\Y_i$ and $\btheta$, is given next.

Let $\widetilde{\ba}_i$ be an expansion point in
a neighborhood of $\ba_i$, $\mathbf{y}_i^+$ be an $\upsilon\times 1$
vector of future measurement of $\mathbf{Y}_i$  (or possibly
missing), $\xp^+_i$ and $\tp^+_i$ be an $\upsilon\times r$ matrix of known
prediction regressors. Then, under the SMSN--NLME model the predictor (or minimum MSE
predictor) of $\mathbf{y^+}_i$ can be approximated as
\begin{eqnarray}
\widetilde{\mathbf{y}}^+_i(\btheta)&=&\E\{\mathbf{y}^+_i|\mathbf{Y}_i,\btheta\}\approx
\widetilde{\bmu}_{2.1}+\frac{\widetilde{\bPsi}_{i22.1}\bupsilon^{(2)}_i}{\sqrt{1+\bupsilon^{(2)\top}_i\widetilde{\bPsi}_{i22.1}\bupsilon^{(2)}_i}}\tau_{-1i},\label{condi}
\end{eqnarray}
where 
$$\widetilde{\bmu}_{2.1}={\eta}(t^+_{ij},d(x^+_{ij},\bbeta,\widetilde{\be}_i))-\widetilde{\bH}^+_i({\widetilde{\ba}}_i-c\bDelta)+\widetilde{\bPsi}^*_{i21}\widetilde{\bPsi}^{*-1}_{i11}\left(\Y_i-{\eta}(t_{ij},d(x_{ij},\bbeta,\widetilde{\be}_i))+\widetilde{\bH}_i({\widetilde{\ba}}_i-c\bDelta)\right),$$
$\widetilde{\bPsi}_{i22.1}=\widetilde{\bPsi}^*_{i22}-\widetilde{\bPsi}^*_{i21}\widetilde{\bPsi}^{*-1}_{i11}\widetilde{\bPsi}^*_{i12},$
$\widetilde{\bPsi}^{*}_{i11}=\widetilde{\bPsi}_i=\widetilde{\bH}_i
\bD \widetilde{\bH}_i^{\top}+\sigma^2 \mathbf{I}_{n_i}$,
$\widetilde{\bPsi}^*_{i12}=\widetilde{\bPsi}^*_{i21}=\widetilde{\bH}^+_i
\bD \widetilde{\bH}_i^{+\top}$,
$\widetilde{\bPsi}^{*}_{i22}=\widetilde{\bH}^+_i \bD
\widetilde{\bH}_i^{+\top}+\sigma^2 \mathbf{I}_{\upsilon}$,
$\widetilde{\bPsi}^{*-1/2}_i\widetilde{\bar{\blambda}}_{i}^*=(\bupsilon^{(1)\top}_i,\bupsilon^{(2)\top}_i)^{\top}$
and
$$\tau_{-1i}=E\left\{U^{-1/2}_i\displaystyle{W_{\Phi}\left(U^{1/2}_i\widetilde{\bupsilon}^{\top}_i(\Y_i-{\eta}(t_{ij},d(x_{ij},\bbeta,\widetilde{\be}_i))+\widetilde{\bH}_i({\widetilde{\ba}}_i-c\bDelta))\right)}|\Y_i\right\},$$
with
$\widetilde{\bupsilon}_i=\displaystyle\frac{{\bupsilon}^{(1)}_i+\widetilde{\bPsi}^{*-1}_{i11}\widetilde{\bPsi}^*_{i12}{\bupsilon}^{(2)}_i}{\sqrt{1+{\bupsilon}^{(2)\top}_i\widetilde{\bPsi}^{*}_{i22.1}{\bupsilon}^{(2)}_i}}.$ The expression for $\tau_{-1i}$ are given in \cite{schumacher2020scale}.

\subsection{Confidence intervals via Bootstrap}\label{sec:bootstrap}

Bootstrap methods are statistical tools used in many statistical problems such as calculus of bias, variance, confidence intervals, sampling distributions of the estimators, among others. For technical details, we refer to \cite{EfronH2016}. In this paper, we use the parametric Bootstrap to calculate a confidence interval for the mode (date of the peak) of COVID-19 deaths by country. Also, we construct confidence bands to the estimated COVID-19 deaths curve along the time. We generate $M$ random samples of length $n$ of the model (eq. \ref{modeleq11}--\ref{modSnmis2}) using the estimated parameters and for each random sample, we fit the proposed model and calculate the statistical mode, the  fitted and predicted curve ($\widehat{\mathbf{Y}}_t$). So, for each country, we will have $Mo_1,..., Mo_M$ estimates of the mode and $\widehat{Y}_{t_1},..., \widehat{Y}_{t_M}$. We calculate the percentiles $\alpha/2$ and $1-\alpha/2$ to construct a confidence interval of level $\alpha$ for the mode and for the fitted curve of the COVID-19 deaths. In this notes, we use $\alpha=5\%$.

To propose the confidence interval for the peak of deaths, we select the two dates such that the $97.5\%$ percentile curve coincides with the highest value of the $2.5\%$ percentile curve. Thus, we guarantee that all curves belonging to the confidence band will peak within this interval. 


\section{Data analysis}\label{s:application}

In order to analyse the COVID-19 data described in Section \ref{s:data}, we propose to fit the SMSN-NLME model given in (\ref{modeleq11}) with the derivative of the generalized logistic model as nonlinear function, which can be written as follows:
\begin{equation}\label{eq:logistfn}
    \eta(t_{ij},\bbeta_{ij})= \frac{\alpha_1 \,\alpha_{3i}\,\alpha_{4}\, \exp\{-\alpha_{3i}t_{ij}\}}{(\alpha_{2i} + \exp\{-\alpha_{3i}\,t_{ij}\})^{\alpha_4+1}},
\end{equation}
where $\alpha_{2i} = \exp\{\beta_2 + b_{2i}\}$, $\alpha_{3i} = \exp\{\beta_3 + b_{3i}\}$, and $\alpha_k = \exp\{\beta_k\}$, for $k=1,4$, with the exponential transformation being used to ensure positiveness of the parameters. Note that this function is similar to the one considered in the univariate approach of \citet{covidlp}, as presented in (\ref{eq:covidlp}), except that in (\ref{eq:logistfn}) random effects are included to enable a multivariate approach and borrow information between the different time series.

For numerical stability, we use the linear transformation $y_{ij} = z_{ij}/k_z$, where $z_{ij}$ is the number of registered deaths at the $i$th country and $j$th day since first death, and $k_z = 33.944$ is chosen to be the sample standard deviation from the Colombia data, which is the smaller one in the observed data.
For model selection, we consider AIC and BIC, as described in Section \ref{subsec:approxML}, for distributions normal (N), skew-normal (SN), student's-t (\emph{t}), and skew-t (ST).
As shown in Table \ref{tab:criteria}, the distribution with lowest AIC and BIC is the ST, which will be used in the analysis hereafter. It is worth noting that the fitted ST distribution has $\nu = 1.568$ (Table \ref{tab:estimates}), evidencing the need for heavy tail models.

\begin{table}[ht]
\caption{Selection criteria for fitting SMSN-NLME models. Bold values indicate the smallest value from each criterion.}
\label{tab:criteria}
\centering
\begin{tabular}{@{}crrr@{}}
\toprule
Distribution & \multicolumn{1}{c}{Loglik} & \multicolumn{1}{c}{AIC} & \multicolumn{1}{c}{BIC} \\ \midrule
N & -2824.9 & 5665.7 & 5704.5 \\
SN & -2824.4 & 5668.8 & 5717.3 \\
\emph{t} & -2360.0 & 4738.1 & 4781.7 \\
ST & \textbf{-2343.3} & \textbf{4708.6} & \textbf{4762.0} \\ \bottomrule
\end{tabular}
\end{table}

\begin{table}[ht]
\centering
\caption{ST-NMLE - ML estimates for parameters, where $\mathbf{F}$ is such that $\mathbf{F}\mathbf{F}=\mathbf{D}$.}\label{tab:estimates}
\begin{tabular}{cccc}
\toprule
Parameter & Estimate & Parameter & Estimate \\\midrule
$\beta_1$ & 14.657 & $F_{11}$ & 0.589 \\
$\beta_2$ & -0.598 & $F_{12}$ & -0.097 \\
$\beta_3$ & -3.127 & $F_{22}$ & 0.721 \\
$\beta_4$ & 2.920 & $\lambda_1$ & -39.985\\
$\sigma^2$ & 3.800 & $\lambda_2$ & 13.915\\
$\nu$ & 1.568  & & \\
\bottomrule
\end{tabular}
\end{table}

\begin{figure}[ht]
    \centering
    \includegraphics[width=\textwidth]{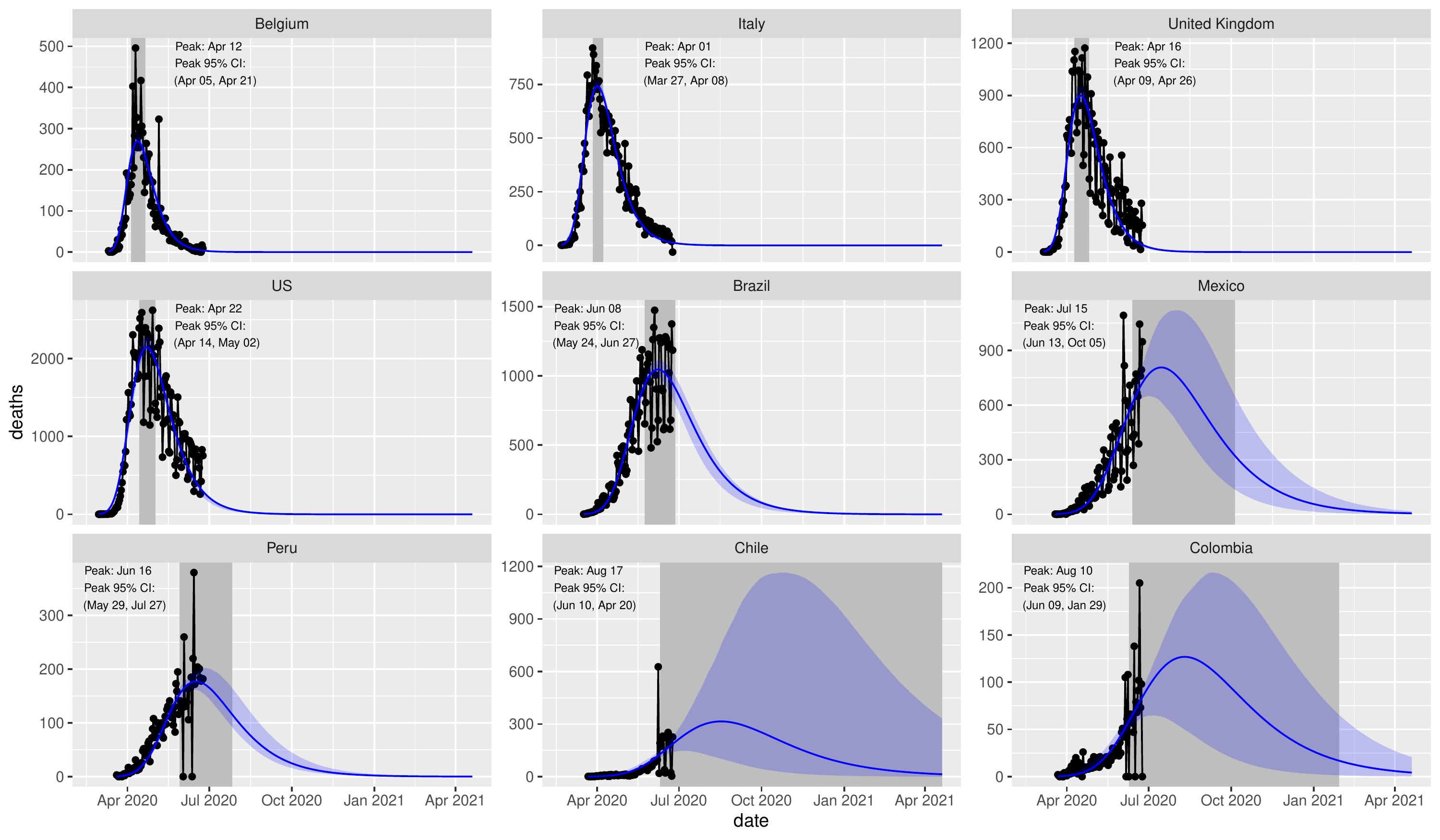}
    \caption{ Fitted and predicted curve for the ST-NLME model (blue line), along with real data (black), and bootstrap 95\% interval estimates for the curve (shaded blue area) and for the peak (shaded grey area), for each country.}
    \label{fig:modelfit}
\end{figure}

To obtain long term evolution estimates, we computed the 300-days-ahead prediction for all countries considered in this study. 
Figure~\ref{fig:modelfit} presents the fitted model, along with the prediction, the real data and confidence intervals information.
To preserve the individual properties observed for each country, the random error was generated conditioned on the mixing scale factor $\widehat{u}_i = \E\{U_i\mid\widehat{\btheta},{\bf y}_i\}$, $i=1,\ldots,9$, which is estimated as a byproduct of the EM algorithm and is responsible to preserve the observed characteristics of the series for each country. Additionally, the 95\% confidence intervals were obtained as described in Subsection~\ref{sec:bootstrap}, where $M=600$ samples were generated, from which the model estimate or prediction for $13$ samples resulted in numerical error and/or non-convergence. For the $587$ remaining samples we performed a 15\% trim of the series to remove those noise series that were not able to replicate the characteristics of the data. To identify such series we used as metric the mean square error (MSE) between the fitted values and observed data and removed those such that the MSE have the highest values. This procedure was performed to prevent convergence problems to affect the interval estimates, resulting in a final of $499$ samples. 
%

From Figure~\ref{fig:modelfit}, we can see that for countries in more advanced stage of the evolution of COVID-19, the confidence intervals are narrow. On the other hand, for countries in early stage such as Chile and Colombia, the uncertainty regarding the disease evolution is much bigger, which is reflected by the wider confidence intervals and very vague information about the peak. 

Table~\ref{tab:forecast} reports predictions for the total number of deaths for each country and up to $30$, $60$, $90$, and $150$-days-ahead of the date from the latest observation considered (2020-06-24), using the fitted ST-NLME model (our selected model), along with its 95\% confidence interval, obtained from the bootstrap results. As can be seen, based on the observed data, the European countries and the US have the deaths by COVID-19 practically controlled, since the death estimates are stable in all future prediction times. Brazil, Mexico and Peru are around its peak (Figure~\ref{tab:forecast}). Their prediction mildly increase with time and they present a little more imprecision as observed by their 95\% confidence interval estimates if compared to the controlled countries. Finally, Chile and Colombia seem to be in the increasing phase. The predicted number of deaths changes drastically between the time intervals with a wide uncertainty.
\begin{table}[ht]\small
\caption{Prediction for $30$, $60$, $90$, and $150$-days-ahead of the total number of deaths using the fitted ST-NLME model. Values inside parenthesis are the 95\% confidence intervals.}\label{tab:forecast}
\begin{center}
\begin{tabular}{@{}ccc@{}}
\toprule
Country & Total expected by 2020-07-25 & Total expected by 2020-08-24 \\ \midrule
Belgium & 10\,296 (9\,675; 11\,206) & 10\,301 (9\,678; 11\,212) \\
Italy & 34\,100 (32\,779; 35\,204) & 34\,128 (32\,798; 35\,230) \\
United Kingdom & 42\,181 (39\,420; 44\,465) & 42\,279 (39\,495; 44\,562) \\
US & 125\,165 (117\,506; 128\,895) & 126\,340 (118\,297; 129\,911) \\
Brazil & 75\,275 (70\,409; 79\,020) & 86\,786 (78\,978; 91\,442) \\
Mexico & 47\,037 (42\,284; 53\,337) & 68\,476 (56\,279; 87\,550) \\
Peru & 12\,664 (11\,795; 14\,170) & 15\,384 (13\,937; 18\,089) \\
Chile & 12\,386 (9\,477; 16\,034) & 21\,640 (13\,024; 35\,669) \\
\vspace{.1cm}
Colombia & 5\,597 (4\,306; 6\,449) & 9\,359 (5\,852; 12\,001) \\ \midrule
Country & Total expected by 2020-09-23 & Total expected by 2020-11-22 \\ \midrule
Belgium & 10\,302 (9\,679; 11\,213) & 10\,302 (9\,679; 11\,213) \\
Italy & 34\,132 (32\,800; 35\,234) & 34\,133 (32\,801; 35\,235) \\
United Kingdom & 42\,295 (39\,506; 44\,577) & 42\,299 (39\,509; 44\,580) \\
US & 126\,627 (118\,467; 130\,158) & 126\,713 (118\,512; 130\,230) \\
Brazil & 91\,894 (82\,288; 96\,861) & 94\,889 (83\,967; 100\,163) \\
Mexico & 83\,630 (64\,215; 113\,683) & 98\,263 (69\,998; 143\,861) \\
Peru & 16\,761 (14\,942; 20\,432) & 17\,692 (15\,551; 22\,243) \\
Chile & 30\,517 (15\,382; 64\,188) & 42\,754 (17\,081; 134\,072) \\
Colombia & 12\,782 (6\,944; 18\,186) & 17\,242 (7\,783; 29\,461)\\
\bottomrule
\end{tabular}
\end{center}
\end{table}

Table~\ref{tab:param} shows the total estimated number of deaths at the end of the pandemic and the estimated peak date by each country. This results corroborates and reinforce the analysis presented in the previous paragraph. Moreover, Table~\ref{tab:param} presents the estimates of the parameters of the logistic dynamic of the model. As it can be seen, $\hat \alpha_4 \gg 1$ showing a large right skewness for the pandemic dynamics. In other words, the increase phase occurs much faster than the decrease phase as observed in many places. This is a clear effect of borrowing strength from curves in different stages. For example, for the curves from the Latin American countries, where observations are mainly on the left side of the peak, estimation of the skewness parameter is unstable due to the lack of information after the peak. Although in similar scale, we can see some variation between $\alpha_{2i}$ and $\alpha_{3i}$, $i=1,\ldots,9$, which are important to capture the unique characteristic of each time series and provide a good fit and meaningful predictions. 
\begin{table}[ht]
\centering
\caption{ST-NMLE - fitted parameters for the generalized logistic curve for the mean. The total estimated number of death (Tot. Est. Death) at the end of the pandemic and the estimated date of the peak (Est. Peak Date) are also presented.}\label{tab:param}
\begin{tabular}{ccccccc}
\toprule
Country &  \multicolumn{1}{c}{$\alpha_1$} &  \multicolumn{1}{c}{$\alpha_{2i}$} &  \multicolumn{1}{c}{$\alpha_{3i}$} &  \multicolumn{1}{c}{$\alpha_4$} & Tot. Est. Death & Est. Peak Date \\\midrule
Belgium & \multirow{9}{*}{78,771,346} & 1.619 & 0.073 & \multirow{9}{*}{18.55} & 10,304 &2020-04-12 \\
Italy & & 1.518 & 0.061 &  & 341,37 &2020-04-01 \\
United Kingdom & & 1.501 & 0.059 &  & 42,303 & 2020-04-16 \\
US & & 1.414 & 0.047 & & 126,726 & 2020-04-22\\
Brazil & & 1.436 & 0.031 & & 95,476 & 2020-06-08\\
Mexico & & 1.429 & 0.022 & & 104,497 & 2020-07-15 \\
Peru & & 1.572 & 0.028 & & 17,922 & 2020-06-16 \\
Chile & & 1.484 & 0.017 & & 51,891 & 2020-08-17 \\
Colombia & & 1.561 & 0.017 & & 20,312 & 2020-08-10\\\bottomrule
\end{tabular}
\end{table}

\section{Conclusion}
\label{s:conc}

This article proposes a robust modeling of COVID-19 deaths based on a NLME model, where the Gaussian distribution of the random terms are replaced by the SMSN class of distributions. We believe that the proposed methods may have a significant impact on COVID-19 research because, in the presence of skewness and heavy tails in the data, appropriate statistical inference for COVID-19 research can lead to more accurate and 
reliable clinical decisions, in addition, our model can be useful to guide some policies that need to be taken by the selected Latin American governments in order to overcome the COVID-19 pandemic. To the best of our knowledge, this is the first attempt in working on such general distributional structure related to COVID-19 deaths data. Our proposed method is quite general and is not limited to the analysis of the selected countries and reported deaths, thus we have made the \texttt{R} codes available for download from \texttt{Github} (\url{https://github.com/fernandalschumacher/NLMECOVID19}), which will encourage other researchers to use NLME models and the SMSN class of
distributions in their studies of other characteristics of COVID-19 data.

Understanding the dynamics of the pandemic and being able to predict its future behavior is of extreme importance. While many countries and consortia try to create an effective vaccine for the disease, the best current alternative is to flatten the contamination curve to guarantee that the health systems are able to provide the necessary care for the population without being overwhelmed. With the use of countries that have controlled the pandemic number of deaths we strongly believe that our methodology can provide more reliable and meaningful prediction that allow policies makers in Latin America to make effective decisions.

Extension of the presented methodology to the number of cases can also be performed, however, because of the large subnotification in the data and due to the different testing capability among countries, this approach is challenging. A possible alternative to overcome this fact is to incorporate into modeling time changing covariates that adequately capture the subnotification dynamics of each country. Another interesting future development and work is to model the cases and deaths by COVID-19 jointly since these series are necessarily correlated. Moreover, the WHO has warned that the coronavirus may never go away and concerns are growing about a second (perhaps a third) wave of infections. Thus, a natural generalization of our method is to consider a finite mixture of SMSN-NLME models \citep{lachos2017finite,zeller2019finite}. An in-depth investigation of such extensions are beyond the scope of the present paper, but certainly an interesting topic for (near) future research.   


\section*{Acknowledgments}
This paper was written while Marcos O. Prates was a visiting professor in the Department of Statistics at the University of Connecticut (UConn). In addition to the support of UConn, the professor would like also to thank the CovidLP group for the discussion about the topic and the Conselho Nacional de Desenvolvimento Científico e Tecnológico (CNPq) for partial financial support. Fernanda L. Schumacher acknowledges the partial support of Coordenação de Aperfeiçoamento de Pessoal de Nível Superior - Brasil (CAPES) - Finance Code 001, and by Conselho Nacional de Desenvolvimento Científico e Tecnológico - Brasil (CNPq).

\bibliographystyle{elsarticle-harv}
\bibliography{biblio}

\begin{thebibliography}{26}
\expandafter\ifx\csname natexlab\endcsname\relax\def\natexlab#1{#1}\fi
\providecommand{\url}[1]{\texttt{#1}}
\providecommand{\href}[2]{#2}
\providecommand{\path}[1]{#1}
\providecommand{\DOIprefix}{doi:}
\providecommand{\ArXivprefix}{arXiv:}
\providecommand{\URLprefix}{URL: }
\providecommand{\Pubmedprefix}{pmid:}
\providecommand{\doi}[1]{\href{http://dx.doi.org/#1}{\path{#1}}}
\providecommand{\Pubmed}[1]{\href{pmid:#1}{\path{#1}}}
\providecommand{\bibinfo}[2]{#2}
\ifx\xfnm\relax \def\xfnm[#1]{\unskip,\space#1}\fi
\bibitem[{Amaro et~al.(2020)Amaro, Dudouet and Orce}]{amaro2020global}
\bibinfo{author}{Amaro, J.E.}, \bibinfo{author}{Dudouet, J.},
  \bibinfo{author}{Orce, J.N.}, \bibinfo{year}{2020}.
\newblock \bibinfo{title}{Global analysis of the covid-19 pandemic using simple
  epidemiological models}.
\newblock \href{http://arxiv.org/abs/2005.06742}{{\tt arXiv:2005.06742}}.
\bibitem[{Branco and Dey(2001)}]{Branco_Dey01}
\bibinfo{author}{Branco, M.D.}, \bibinfo{author}{Dey, D.K.},
  \bibinfo{year}{2001}.
\newblock \bibinfo{title}{A general class of multivariate skew-elliptical
  distributions}.
\newblock \bibinfo{journal}{Journal of Multivariate Analysis}
  \bibinfo{volume}{79}, \bibinfo{pages}{99--113}.
\bibitem[{Cleveland(1979)}]{cleveland1979robust}
\bibinfo{author}{Cleveland, W.S.}, \bibinfo{year}{1979}.
\newblock \bibinfo{title}{Robust locally weighted regression and smoothing
  scatterplots}.
\newblock \bibinfo{journal}{Journal of the American Statistical Association}
  \bibinfo{volume}{74}, \bibinfo{pages}{829--836}.
\bibitem[{{CovidLP Team}(2020)}]{covidlp}
\bibinfo{author}{{CovidLP Team}}, \bibinfo{year}{2020}.
\newblock \bibinfo{title}{CovidLP: short and long term prediction for
  COVID-19}.
\newblock \bibinfo{organization}{Departamento de Estat\'istica}.
  \bibinfo{address}{UFMG, Brazil}.
\newblock \URLprefix \url{http://est.ufmg.br/covidlp/home/en/}.
\bibitem[{{De Castro}(2020)}]{arm2020sir}
\bibinfo{author}{{De Castro}, C.A.}, \bibinfo{year}{2020}.
\newblock \bibinfo{title}{Sir model for {COVID-19} calibrated with existing
  data and projected for colombia}.
\newblock \href{http://arxiv.org/abs/2003.11230}{{\tt arXiv:2003.11230}}.
\bibitem[{Dempster et~al.(1977)Dempster, Laird and Rubin}]{Dempster77}
\bibinfo{author}{Dempster, A.}, \bibinfo{author}{Laird, N.},
  \bibinfo{author}{Rubin, D.}, \bibinfo{year}{1977}.
\newblock \bibinfo{title}{Maximum likelihood from incomplete data via the {EM}
  algorithm}.
\newblock \bibinfo{journal}{Journal of the Royal Statistical Society, Series
  B,} \bibinfo{volume}{39}, \bibinfo{pages}{1--38}.
\bibitem[{Efron and Hastie(2016)}]{EfronH2016}
\bibinfo{author}{Efron, B.}, \bibinfo{author}{Hastie, T.},
  \bibinfo{year}{2016}.
\newblock \bibinfo{title}{Computer age statistical inference}.
\newblock \bibinfo{publisher}{Cambridge University Press},
  \bibinfo{address}{Cambridge}.
\bibitem[{{JF Salvando Todos Team}(2020)}]{JFsalvandotodos}
\bibinfo{author}{{JF Salvando Todos Team}}, \bibinfo{year}{2020}.
\newblock \bibinfo{title}{Plataforma Estatística JF para COVID-19}.
\newblock \bibinfo{organization}{Departamento de Estat\'istica}.
  \bibinfo{address}{UFJF, Brazil}.
\newblock \URLprefix \url{http://jfsalvandotodos.ufjf.br}.
\bibitem[{Lachos et~al.(2010)Lachos, Ghosh and
  Arellano-Valle}]{Lachos_Ghosh_Arellano_2009}
\bibinfo{author}{Lachos, V.H.}, \bibinfo{author}{Ghosh, P.},
  \bibinfo{author}{Arellano-Valle, R.B.}, \bibinfo{year}{2010}.
\newblock \bibinfo{title}{Likelihood based inference for skew--normal
  independent linear mixed models}.
\newblock \bibinfo{journal}{Statistica Sinica} \bibinfo{volume}{20},
  \bibinfo{pages}{303--322}.
\bibitem[{Lachos et~al.(2017)Lachos, Moreno, Chen and
  Cabral}]{lachos2017finite}
\bibinfo{author}{Lachos, V.H.}, \bibinfo{author}{Moreno, E.J.L.},
  \bibinfo{author}{Chen, K.}, \bibinfo{author}{Cabral, C.R.B.},
  \bibinfo{year}{2017}.
\newblock \bibinfo{title}{Finite mixture modeling of censored data using the
  multivariate student-t distribution}.
\newblock \bibinfo{journal}{Journal of Multivariate Analysis}
  \bibinfo{volume}{159}, \bibinfo{pages}{151--167}.
\bibitem[{Lange and Sinsheimer(1993)}]{Lange93}
\bibinfo{author}{Lange, K.L.}, \bibinfo{author}{Sinsheimer, J.S.},
  \bibinfo{year}{1993}.
\newblock \bibinfo{title}{Normal/independent distributions and their
  applications in robust regression}.
\newblock \bibinfo{journal}{J. Comput. Graph. Stat} \bibinfo{volume}{2},
  \bibinfo{pages}{175--198}.
\bibitem[{Lindstrom and Bates(1990)}]{Lindstrombates90}
\bibinfo{author}{Lindstrom, M.}, \bibinfo{author}{Bates, D.},
  \bibinfo{year}{1990}.
\newblock \bibinfo{title}{Nonlinear mixed-effects models for repeated-measures
  data}.
\newblock \bibinfo{journal}{Biometrics} \bibinfo{volume}{46},
  \bibinfo{pages}{673--687}.
\bibitem[{Maugeri et~al.(2020)Maugeri, Barchitta, Battiato and
  Agodi}]{maugeri2020estimation}
\bibinfo{author}{Maugeri, A.}, \bibinfo{author}{Barchitta, M.},
  \bibinfo{author}{Battiato, S.}, \bibinfo{author}{Agodi, A.},
  \bibinfo{year}{2020}.
\newblock \bibinfo{title}{Estimation of unreported novel coronavirus
  (sars-cov-2) infections from reported deaths: A
  susceptible--exposed--infectious--recovered--dead model}.
\newblock \bibinfo{journal}{Journal of Clinical Medicine} \bibinfo{volume}{9},
  \bibinfo{pages}{1350}.
\bibitem[{Mellan et~al.(08-05-2020)Mellan, Hoeltgebaum, Mishraet and
  {al.}}]{ImperialColledge}
\bibinfo{author}{Mellan, T.A.}, \bibinfo{author}{Hoeltgebaum, H.H.},
  \bibinfo{author}{Mishraet, S.}, \bibinfo{author}{{al.}},
  \bibinfo{year}{08-05-2020}.
\newblock \bibinfo{title}{Estimating covid-19 cases and reproduction number in
  brazil}.
\newblock \bibinfo{journal}{Imperial College London}
  \DOIprefix\doi{https://doi.org/10.25561/78872}.
\bibitem[{{R Core Team}(2019)}]{rmanual}
\bibinfo{author}{{R Core Team}}, \bibinfo{year}{2019}.
\newblock \bibinfo{title}{R: A Language and Environment for Statistical
  Computing}.
\newblock \bibinfo{organization}{R Foundation for Statistical Computing}.
  \bibinfo{address}{Vienna, Austria}.
\newblock \URLprefix \url{https://www.R-project.org/}.
\bibitem[{Ribeiro and Bernardes(2020)}]{ribeiro2020underreporting}
\bibinfo{author}{Ribeiro, L.C.}, \bibinfo{author}{Bernardes, A.T.},
  \bibinfo{year}{2020}.
\newblock \bibinfo{title}{{Estimate of underreporting of COVID-19 in Brazil by
  Acute Respiratory Syndrome hospitalization reports}}.
\newblock \bibinfo{type}{Notas Técnicas Cedeplar-UFMG}. Cedeplar, Universidade
  Federal de Minas Gerais.
\newblock \URLprefix \url{https://ideas.repec.org/p/cdp/tecnot/tn010.html}.
\bibitem[{Rivera-Rodriguez and
  Urdinola(2020)}]{Rivera-Rodriguez2020.04.14.20065466}
\bibinfo{author}{Rivera-Rodriguez, C.}, \bibinfo{author}{Urdinola, B.P.},
  \bibinfo{year}{2020}.
\newblock \bibinfo{title}{Predicting hospital demand during the covid-19
  outbreak in bogota, colombia}.
\newblock \bibinfo{journal}{medRxiv} \URLprefix
  \url{https://www.medrxiv.org/content/early/2020/06/24/2020.04.14.20065466},
  \DOIprefix\doi{10.1101/2020.04.14.20065466}.
\bibitem[{Schumacher et~al.(2020a)Schumacher, Dey and
  Lachos}]{schumacher2020approximate}
\bibinfo{author}{Schumacher, F.L.}, \bibinfo{author}{Dey, D.K.},
  \bibinfo{author}{Lachos, V.H.}, \bibinfo{year}{2020}a.
\newblock \bibinfo{title}{Approximate inferences for nonlinear mixed effects
  models with scale mixture of skew-normal distributions}.
\newblock \href{http://arxiv.org/abs/2007.15086}{{\tt arXiv:2007.15086}}.
\bibitem[{Schumacher et~al.(2020b)Schumacher, Lachos and
  Matos}]{schumacher2020scale}
\bibinfo{author}{Schumacher, F.L.}, \bibinfo{author}{Lachos, V.H.},
  \bibinfo{author}{Matos, L.A.}, \bibinfo{year}{2020}b.
\newblock \bibinfo{title}{Scale mixtures of skew-normal linear mixed models
  with within-subject serial dependence}.
\newblock \href{http://arxiv.org/abs/2002.01040}{{\tt arXiv:2002.01040}}.
\bibitem[{Schumacher et~al.(2020c)Schumacher, Matos and
  Lachos}]{skewlmm-manual}
\bibinfo{author}{Schumacher, F.L.}, \bibinfo{author}{Matos, L.A.},
  \bibinfo{author}{Lachos, V.H.}, \bibinfo{year}{2020}c.
\newblock \bibinfo{title}{skewlmm: Scale mixtures of skew-normal linear mixed
  models}.
\newblock \URLprefix \url{https://CRAN.R-project.org/package=skewlmm}.
  \bibinfo{note}{r package version 0.2.0}.
\bibitem[{Torrealba-Rodriguez et~al.(2020)Torrealba-Rodriguez,
  Conde-Guti\'{e}rrez and Hern\'{a}ndez-Javier}]{TorrealbaCH2020}
\bibinfo{author}{Torrealba-Rodriguez, O.},
  \bibinfo{author}{Conde-Guti\'{e}rrez, R.A.},
  \bibinfo{author}{Hern\'{a}ndez-Javier, A.L.}, \bibinfo{year}{2020}.
\newblock \bibinfo{title}{Modeling and prediction of covid-19 in mexico
  applying mathematical and computational models}.
\newblock \bibinfo{journal}{Chaos, Solitons and Fractals}
  \bibinfo{volume}{138}, \bibinfo{pages}{109946}.
\bibitem[{Tsallis and Tirnakli(2020)}]{TsallisT2020}
\bibinfo{author}{Tsallis, C.}, \bibinfo{author}{Tirnakli, U.},
  \bibinfo{year}{2020}.
\newblock \bibinfo{title}{Predicting covid-19 peaks around the world}.
\newblock \bibinfo{journal}{Frontiers in Physics} \bibinfo{volume}{8},
  \bibinfo{pages}{217}.
\newblock \URLprefix
  \url{https://www.frontiersin.org/article/10.3389/fphy.2020.00217},
  \DOIprefix\doi{10.3389/fphy.2020.00217}.
\bibitem[{Wit et~al.(2012)Wit, Heuvel and Romeijn}]{wit2012all}
\bibinfo{author}{Wit, E.}, \bibinfo{author}{Heuvel, E.v.d.},
  \bibinfo{author}{Romeijn, J.W.}, \bibinfo{year}{2012}.
\newblock \bibinfo{title}{"all models are wrong...": an introduction to model
  uncertainty}.
\newblock \bibinfo{journal}{Statistica Neerlandica} \bibinfo{volume}{66},
  \bibinfo{pages}{217--236}.
\bibitem[{Worldometers.info(2020)}]{Worldometers}
\bibinfo{author}{Worldometers.info}, \bibinfo{year}{2020}.
\newblock \bibinfo{title}{Covid-19 coronavirus pandemic}.
\newblock
  \bibinfo{howpublished}{\url{https://www.worldometers.info/coronavirus/\#countries}}.
\newblock \bibinfo{note}{Accessed: 2020-06-26}.
\bibitem[{Wu(2010)}]{Wu2010}
\bibinfo{author}{Wu, L.}, \bibinfo{year}{2010}.
\newblock \bibinfo{title}{Mixed Effects Models for Complex Data}.
\newblock \bibinfo{publisher}{Chapman \& Hall/CRC}, \bibinfo{address}{Boca
  Raton, FL}.
\bibitem[{Zeller et~al.(2019)Zeller, Cabral, Lachos and
  Benites}]{zeller2019finite}
\bibinfo{author}{Zeller, C.B.}, \bibinfo{author}{Cabral, C.R.B.},
  \bibinfo{author}{Lachos, V.H.}, \bibinfo{author}{Benites, L.},
  \bibinfo{year}{2019}.
\newblock \bibinfo{title}{Finite mixture of regression models for censored data
  based on scale mixtures of normal distributions}.
\newblock \bibinfo{journal}{Advances in Data Analysis and Classification}
  \bibinfo{volume}{13}, \bibinfo{pages}{89--116}.

\end{thebibliography}

\end{document}